\documentclass[structabstract]{aa}  
\usepackage{graphicx}
\usepackage{txfonts}
\usepackage{natbib}
\bibpunct{(}{)}{;}{a}{}{,}
\begin{document}
   \title{The frequency of giant planets around metal-poor stars\thanks{The data presented herein are based on observations collected at the La Silla Parana
Observatory, ESO (Chile) with the HARPS spectrograph at
the 3.6-m telescope (ESO runs ID 72.C-0488, 082.C-0212, and
085.C-0063) and at the
W.M. Keck Observatory that is operated as a scientific partnership among the California Institute of Technology, the University of California and the National Aeronautics and Space
Administration. This Observatory was made possible by the
generous financial support of the W.M. Keck Foundation.}\thanks{Table 1 is only available in electronic form
at the CDS via anonymous ftp to cdsarc.u-strasbg.fr (130.79.128.5)
or via http://cdsweb.u-strasbg.fr/cgi-bin/qcat?J/A+A/}}

   \subtitle{}

   \author{A. Mortier\inst{1} \and
           N.C.Santos\inst{1,2}\and
           A. Sozzetti\inst{3}\and
	   M. Mayor\inst{4}\and
	   D. Latham\inst{5}\and
	   X. Bonfils\inst{6}\and
	   S. Udry\inst{4}
          }

   \institute{Centro de Astrof\'{\i}sica, Universidade do Porto, Rua das Estrelas, 4150-762 Porto, Portugal\\
              \email{amortier@astro.up.pt}
	      \and
	      Departamento de F\'{\i}sica e Astronomia, Faculdade de Ci\^encias, Universidade do Porto, Portugal
	      \and
	      INAF-Osservatorio Astronomico di Torino, strada dell'Osservatorio 20, 10025 Pino Torinese, Italy
	      \and
	      Observatoire de Genève, Universit\'e de Genève, 51 Ch. des Maillettes, 1290 Sauverny, Switzerland
	      \and
	      Harvard-Smithsonian Center for Astrophysics, 60 Garden Street, Cambridge, MA 02138, USA
	      \and
	      UJF-Grenoble 1 / CNRS-INSU, Institut de Planètologie et d’Astrophysique de Grenoble (IPAG) UMR 5274, Grenoble, F-38041, France
	      }

   \date{Received ...; Accepted ...}

 
  \abstract
   {The discovery of about 700 extrasolar planets, so far, has lead to the first statistics concerning extrasolar planets. The presence of giant planets seems to depend on stellar metallicity and mass. For example, they are more frequent around metal-rich stars, with an exponential increase in planet occurrence rates with metallicity.}
   {We analyzed two samples of metal-poor stars ($-2.0\leq$ [Fe/H] $\leq 0.0$) to see if giant planets are indeed rare around these objects. Radial velocity datasets were obtained with two different spectrographs (HARPS and HIRES). Detection limits for these data, expressed in minimum planetary mass and period, are calculated. These produce trustworthy numbers for the planet frequency.}
   {A general Lomb Scargle (GLS) periodogram analysis was used together with a bootstrapping method to produce the detection limits. Planet frequencies were calculated based on a binomial distribution function within metallicity bins.}
   {Almost all hot Jupiters and most giant planets should have been found in these data. Hot Jupiters around metal-poor stars have a frequency lower than $1.0\%$  at one sigma. Giant planets with periods up to $1800$ days, however, have a higher frequency of $f_p=2.63^{+2.5}_{-0.8}\%$. Taking into account the different metallicities of the stars, we show that giant planets appear to be very frequent ($f_p=4.48^{+4.04}_{-1.38}\%$) around stars with [Fe/H] $>-0.7$, while they are rare around stars with [Fe/H] $\leq -0.7$ ($\leq 2.36\%$ at one sigma).}
   {Giant planet frequency is indeed a strong function of metallicity, even in the low-metallicity tail. However, the frequencies are most likely higher than previously thought.}

   \keywords{Planetary systems -- Techniques: radial velocities
               }

   \authorrunning{Mortier, A. et al.}
   \maketitle
%

\section{Introduction}

Since the discovery of the first extrasolar planet in 1995 \citep[51 Peg b, ][]{Mayor95}, the search for extrasolar planetary systems accelerated. Today, around 750 planets are announced. Most of them were detected using the radial velocity technique. Although 750 is a relatively high number, the theory of planet formation and evolution is still under debate \citep{Pol96,May02,Mor09}. The situation is particularly difficult for giant planet formation. Currently, there are two proposed models: core-accretion \citep[e.g.][]{Pol96,Rice03,Ali04}, where gas from the protoplanetary disk is accreted around a previously formed rocky/icy core, and the disk instability model \citep[e.g.][]{Boss97,May02}, where a planet is formed because of a direct gravitational instability in the protoplanetary disk, in the same way as stars form from interstellar clouds. A helpful overview of both models is given by \citet{Mat07}.

One of the main advantages of the instability model is the timescale that is needed to form the planets. Early results suggested that the slow accretion phase (about 10Myr) in the core-accretion model may take longer than the lifetime of a T Tauri disk \citep{Pol96}. In that sense, giant planets could not form within the core-accretion model. However, more recent results suggest that this may not be a real problem. In fact, it has been shown that the process can, for example, be accelerated by including disk-induced orbital migration \citep{Ali04,Mor09}.

Theories of migration became more important with the discovery of 51 Peg b and other {\it hot Jupiters}. These close-in giant planets are highly unlikely to have formed in-situ. Interestingly, however, disk-induced migration does not necessarily provide the correct explanation for all these hot Jupiters. Other ideas have been put forward \citep[e.g.][]{Tri10,MJ11,Soc12}, which include more ``violent" migration mechanisms. Discoveries of giant planets on wide orbits of tens to hundreds AU \citep{Mar08,Lag10} also raise questions. Overall there is still space to support that the disk-instability model may be at work, at least to explain part of the detected planets \citep{Vor10,Boss11}.

Additional clues about this problem come from the analysis of planet-host stars. The presence of a planet seems to depend on several stellar properties, such as mass and metallicity \citep{Udry07}. Concerning metallicity, it has been well-established that more metal-rich stars have a higher probability of harboring a giant planet than their lower metallicity counterparts \citep{Gon97,San01,San04,Fis05,Udry07,Soz09,Sou11b}. The occurrence rate even increases dramatically with increasing metallicity. Current numbers, based on the CORALIE and HARPS samples, suggest that around $25\%$ of the stars with twice the metal content of our Sun are orbited by a giant planet. This number decreases to $\sim5\%$ for solar-metallicity objects \citep{Sou11b,Mayor11}. A similar trend was also obtained by previous results \citep[e.g.][]{San04,Fis05,John10}.
Curiously, no such trend is observed for the lower mass planets \citep{Udry06,Sou08,Mayor11}. The Neptune-mass planets found so far seem to have a rather flat metallicity distribution \citep{Sou08,Sou11b,Mayor11}.

This observed metallicity correlation favors the core-accretion model for the formation of giant planets \citep{Ida04,Udry07,Mor12} because the higher the grain content of the disk, the easier it is to build the cores that will later accrete gas. According to the disk-instability model, the presence of planets would not be strongly dependent on stellar metallicity \citep{Boss02}.

Understanding the frequency of different types of planets around stars of different mass and metallicity is thus providing clues about the processes of planet formation and evolution. This has inspired the construction of specific samples to search for planets around different types of stars \citep[e.g.][]{San07,Sat08,Soz09}. Statistics of these samples will help in understanding the formation processes and constrain the models.

We present an analysis of two metal-poor samples that were designed for planet-finding purposes. In Section \ref{Dat}, an overview is given of the samples and their data. Section \ref{Lim} reports on the detection limits of these samples. Planet frequencies are calculated in Section \ref{Freq}. Finally, conclusions are made in Section \ref{Con}, together with a discussion.


\section{Data}\label{Dat}

Radial velocity measurements from two different samples of metal-poor, solar-type stars were used in this paper. 

\subsection{The HARPS sample}

\begin{figure}[t!]
\begin{center}
\includegraphics[width=4.4cm]{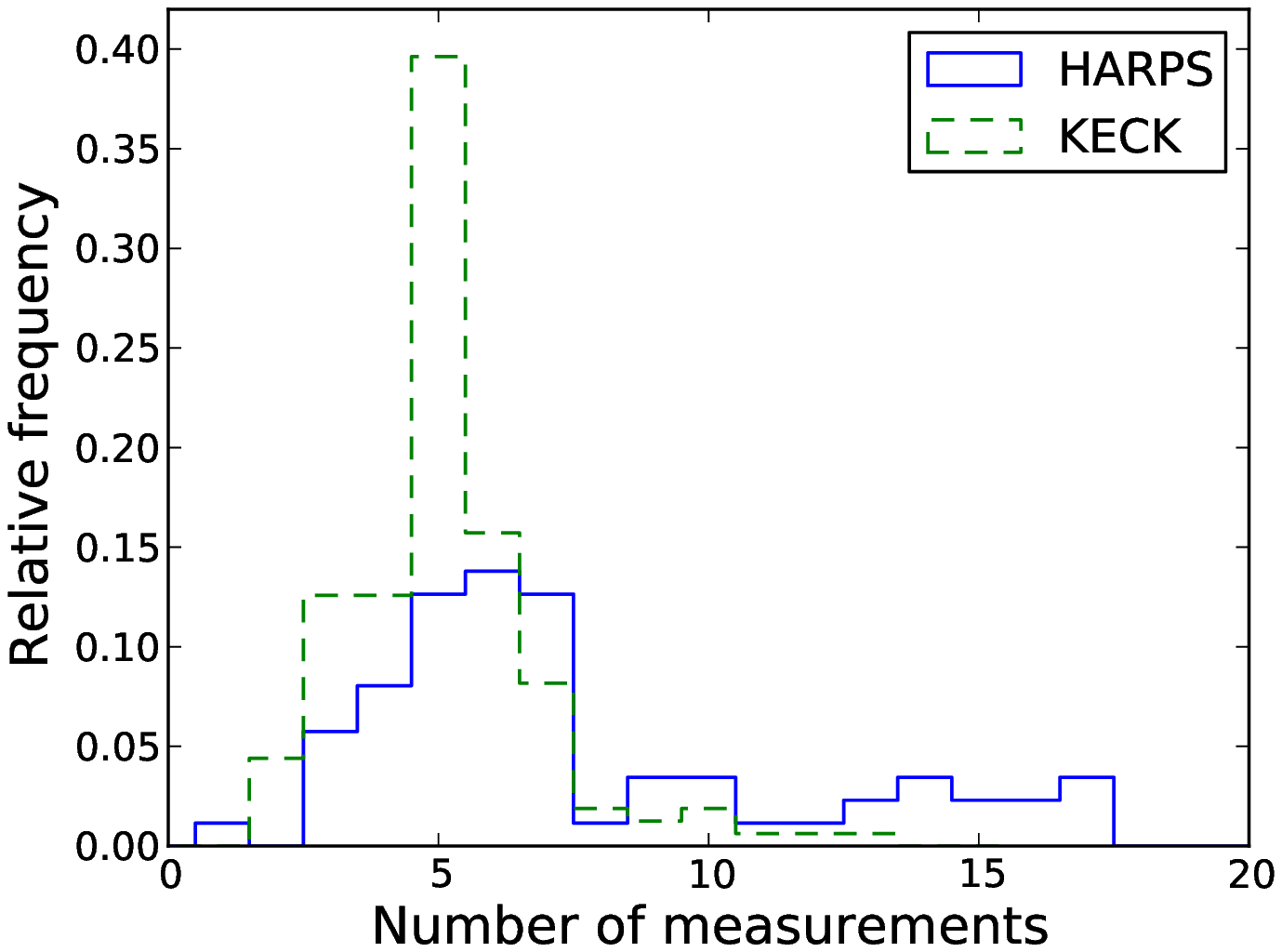}
\includegraphics[width=4.4cm]{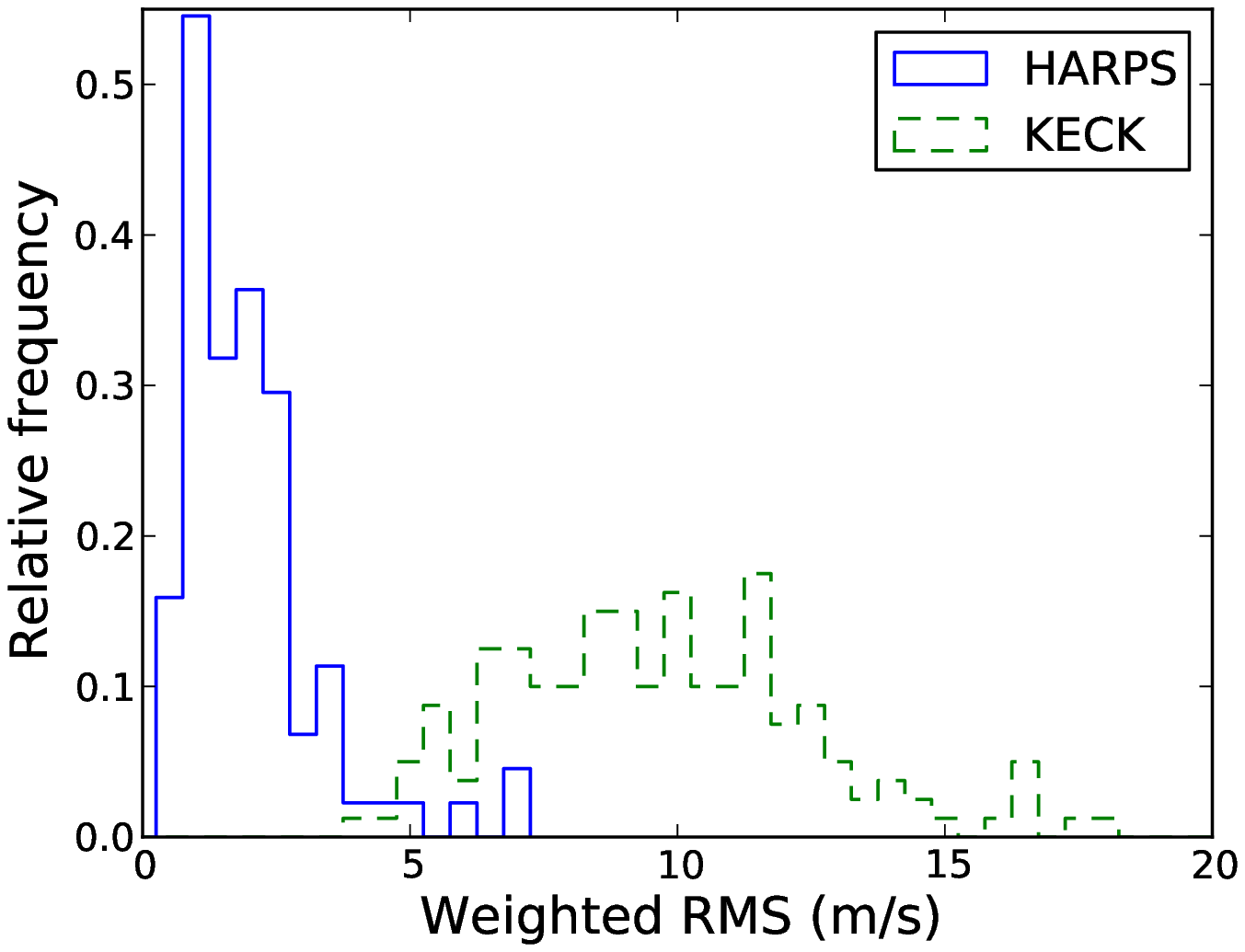}\\
\includegraphics[width=4.4cm]{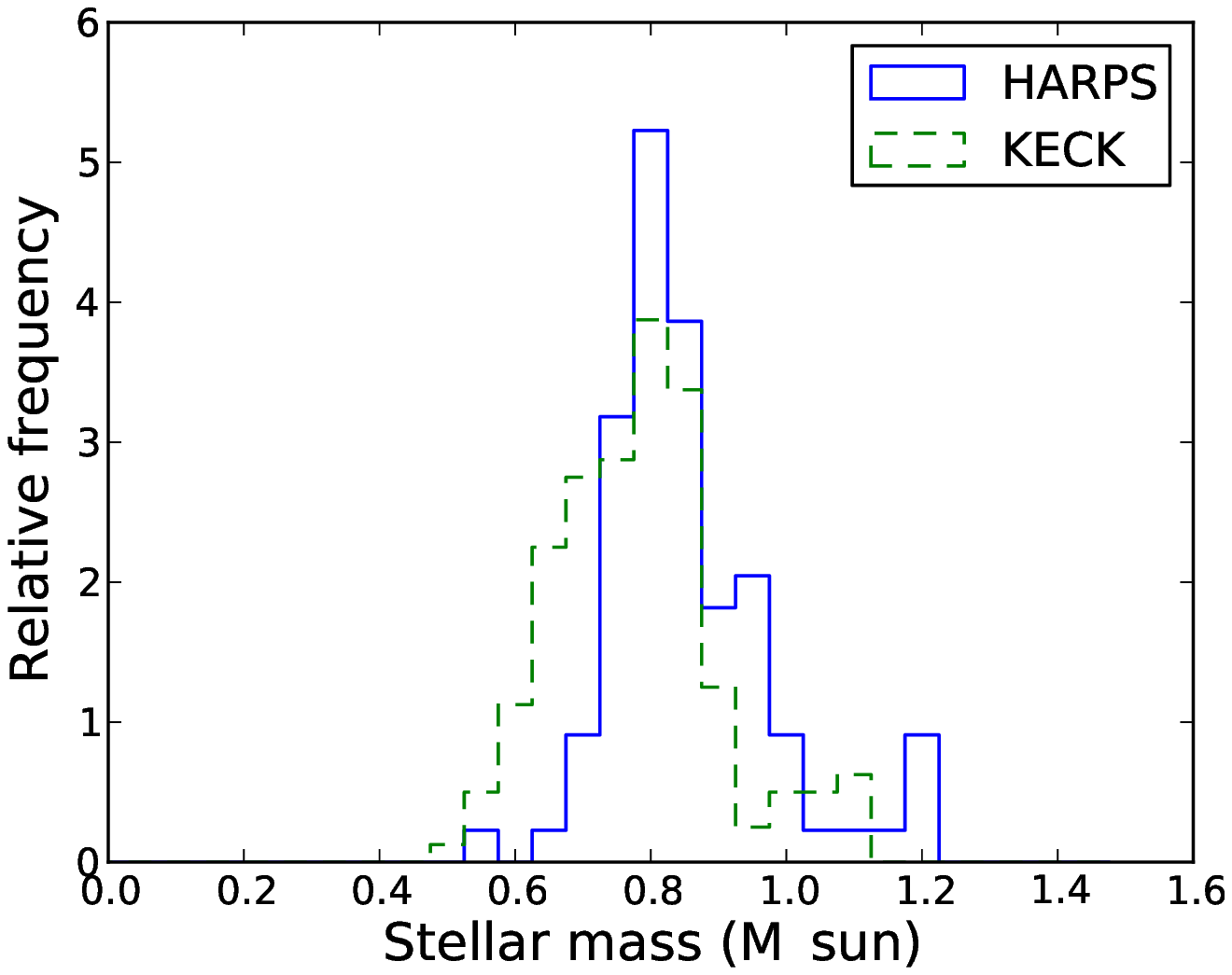}
\includegraphics[width=4.4cm]{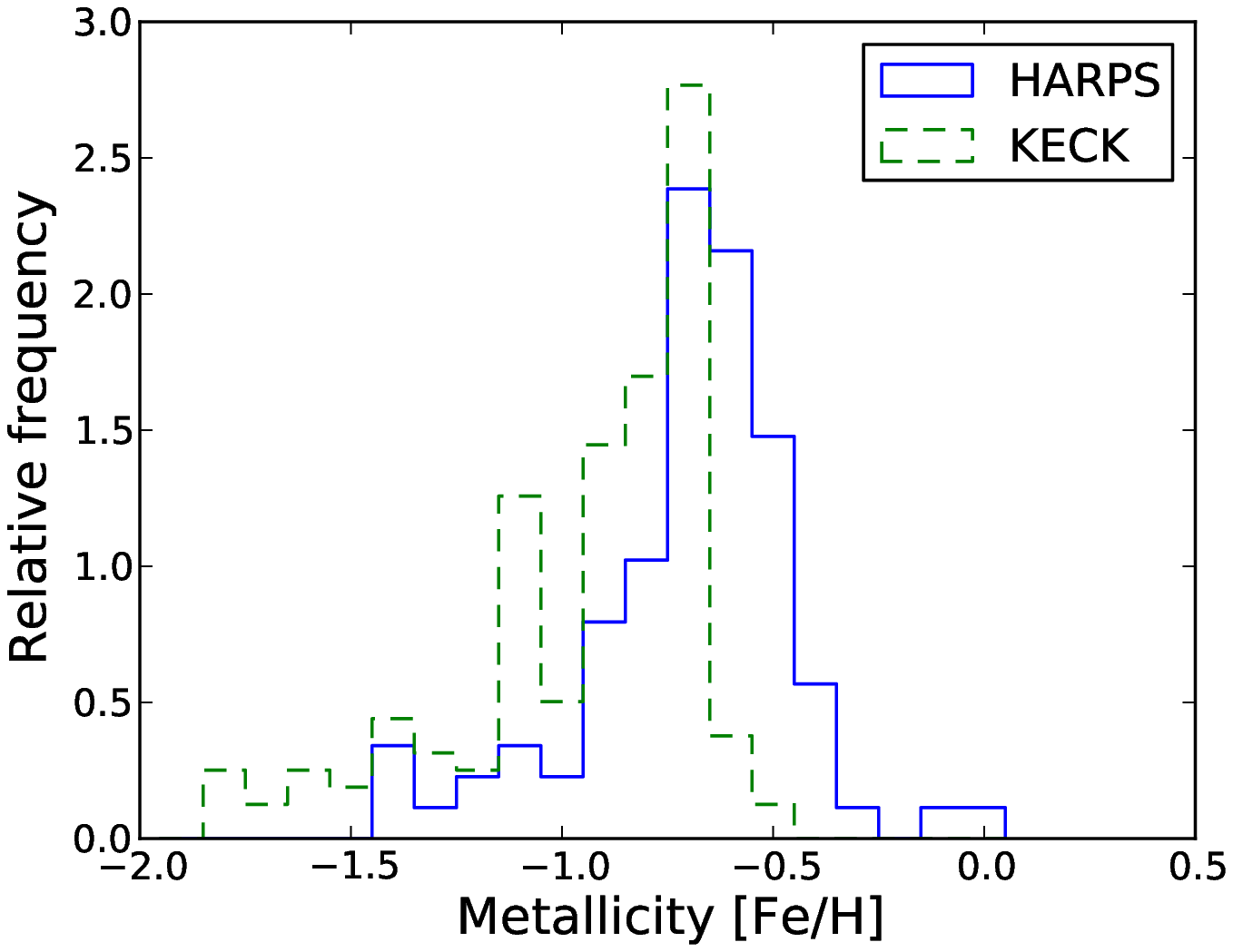}
\caption{Relative histogram of the number of measurements (top left panel), rms noise (top right panel), stellar mass (bottom left panel) and metallicity (bottom right panel) in the two datasets. The blue line represents the HARPS sample, the green line the KECK-HIRES sample. Both graphs in the top panel are cut at 20 for better visibility. The top left panel has 19 stars from the HARPS sample higher then 20. The top right panel has 1 star from the KECK sample higher then 20.}
\label{FigData}
\end{center}
\end{figure}

\citet{San11} reported on the first sample. They observed 104 metal-poor or mild metal-poor solar-type stars with the HARPS spectrograph \citep{Mayor03}. The objects were observed from October 2003 till July 2010.
Based on the catalog of \citet{Nord04}, all late-F, G and K stars south of $+10^{\circ}$ declination with a V magnitude brighter than 12 were chosen. After discarding spectroscopic binaries, giant stars and active stars, Santos and coworkers only recovered the 104 stars with an estimated photometric [Fe/H] between $-0.5$ and $-1.5$. After the observations, 16 stars were also discarded because they were unsuitable targets for planet-finding purposes (binarity, activity, high rotation).
Most of the stars in the final sample of 88 stars have five or more measurements with an rms of $\sim1-2.5$ m\,s$^{-1}$ as shown in Fig. \ref{FigData} and Table \ref{TabData}. In the bottom panels of Fig. \ref{FigData}, the mass and metallicity distribution is shown. Values were taken from \citet{Sou11a}. The (spectroscopic) metallicities differ slightly from the photometric estimate that was initially used, including a few outliers. All values are listed in Table \ref{TabData}.

Three planetary mass companions were found in this sample. They are orbiting \object{HD181720}, \object{HD190984} \citep{San10}, and \object{HD171028} \citep{San07}. All three are giant planets in long-period orbits. 
A fourth planet candidate, orbiting \object{HD107094}, was announced in \citet{San11}. With a 4.5 $M_{Jup}$ minimum mass and a 1870 day period, it would again be a giant planet in a long-period orbit. However, this planet could not be fully confirmed yet.

\subsection{The KECK-HIRES sample}

\citet{Soz09} reported on the second sample. They observed 160 metal-poor solar-type stars with the HIRES spectrograph on the Keck 1 telescope at Mauna Kea in Hawaii \citep{Vogt94}. The objects were all observed at least twice over a timespan of three years (2003 - 2006). This sample of stars was drawn from the Carney-Latham and Ryan samples \citep[e.g.][]{Car94,Ryan89}. Additional criteria were applied, and in the end, Sozzetti and collaborators chose the 160 stars with a V magnitude brighter than 12, an effective temperature T$_{eff}$ lower than 6000K and a metallicity $[Fe/H]$ between $-0.4$ and $-1.8$. All stars are situated north of $-25^{\circ}$ declination. Most stars in this sample have 4 to 10 measurements with a rms of $\sim 9$ m\,s$^{-1}$ as seen in Fig. \ref{FigData} and Table \ref{TabData}. Photon noise is the main contributor to these higher rms values. The mass and metallicity values, seen in this figure, are taken from \citet{Soz09}. All values are listed in Table \ref{TabData}. Typical uncertainties on T$_{eff}$ , [Fe/H], and M$_{\ast}$ are 100 K, 0.1 dex, and 0.1 M$_{\odot}$ respectively.

No planetary signals were found in this sample.

\begin{table*}
\caption{Relevant values for the targets in the two samples. The complete table is provided in electronic form only. }
\label{TabData}     
\centering                       
\begin{tabular}{llllllll}
\hline\hline 
Star & n & Mean rms & Timespan & T$_{eff}$ & [Fe/H] & M$_{\ast}$ & Reference\\
 &  & [m\,s$^{-1}$] & [days] & [K] &  & [M$_{\odot}$] & \\                
\hline
HD123517 & 9 & 3.02 & 1596 & 6082$\pm$29 & 0.09$\pm$0.02 & 1.21$\pm$0.08 & (1)\\
HD124785 & 17 & 2.01 & 1518 & 5867$\pm$21 & -0.56$\pm$0.01 & 0.87$\pm$0.02 & (1)\\
HD126681 & 13 & 2.23 & 1964 & 5570$\pm$34 & -1.15$\pm$0.03 & 0.71$\pm$0.02 & (1)\\
G15-7 & 6 & 9.27 & 890 & 5280 & -0.88 & 0.74 & (2)\\
G151-10 & 7 & 10.91 & 891 & 5287 & -0.70 & 0.76 & (2)\\
G157-93 & 3 & 4.43 & 115 & 5409 & -0.78 & 0.78 & (2)\\
... & ... & ... & ... & ... & ... & ... & ...\\
\hline                       
\end{tabular}
\tablebib{
(1) ̃\citet{Sou11a}; (2) \citet{Soz09}.
}
\end{table*}

\subsection{The combined sample}

Fourteen (14) stars have measurements in both samples. This amounts to a complete sample of 234 metal-poor solar-type stars. The two samples use different naming for the stars (see Table \ref{TabCom}). In this paper, the naming from the HARPS sample will be used for these 14 stars.

\begin{table}
\caption{Names and number of measurements of stars that are present in the two samples.}
\label{TabCom}     
\centering                       
\begin{tabular}{llll}
\hline\hline                 
name HARPS & $n_H$ & name KECK & $n_K$ \\
\hline
\object{BDp062932} & 4 & \object{G66-22} & 4 \\
\object{BDp083095} & 3 & \object{G16-13} & 5 \\
\object{HD104800} & 6 & \object{G11-36} & 4 \\
\object{HD111515} & 5 & \object{G14-5} & 7 \\
\object{HD126681} & 13 & \object{HD126681} & 6 \\
\object{HD131653} & 4 & \object{G151-10} & 7 \\
\object{HD134440} & 10 & \object{HD134440} & 6 \\
\object{HD148211} & 31 & \object{HD148211} & 3 \\
\object{HD148816} & 6 & \object{HD148816} & 10 \\
\object{HD193901} & 3 & \object{HD193901} & 5 \\
\object{HD196892} & 3 & \object{HD196892} & 5 \\
\object{HD215257} & 37 & \object{G27-44} & 8 \\
\object{HD22879} & 36 & \object{G80-15} & 5 \\
\object{HD88725} & 22 & \object{G44-6} & 5 \\
\hline                       
\end{tabular}
\end{table}

Combining the measurements of these stars provides more data to look for possible planetary signals. The data of two different telescopes were combined by subtracting the mean of the data from each set in the overlapping time-interval. A general Lomb-Scargle (GLS) periodogram analysis (see Section \ref{Lim}) was then performed on the 14 stars. No significant peaks were found in these periodograms.

%

\section{Detection limits}

\subsection{Methodology}\label{Lim}

In the literature, different authors used two main approaches to find detection limits in RV data. One is based on $\chi^2$- and $F$-tests \citep[e.g.][]{Lag09,Soz09}, another is based on a periodogram analysis \citep[e.g.][]{Cum99,Cum04,Endl01,Nar05}.  In this paper, the second approach was chosen because we consider that we have enough measurements for a reliable periodogram analysis
(see below for a comparison of the methods).

\begin{figure}[t!]
\begin{center}
\includegraphics[width=6.7cm,angle=270]{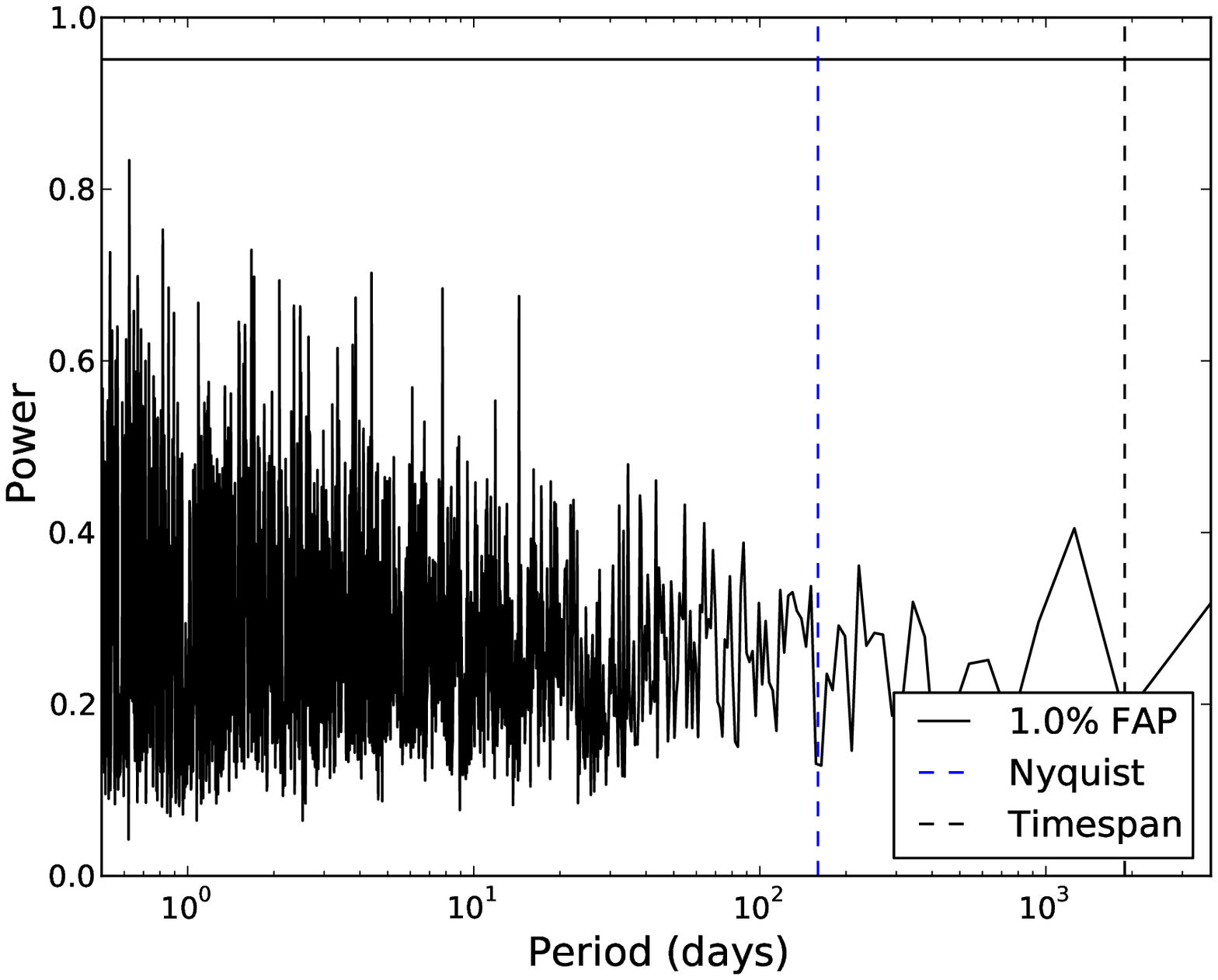}\\
\includegraphics[width=6.7cm,angle=270]{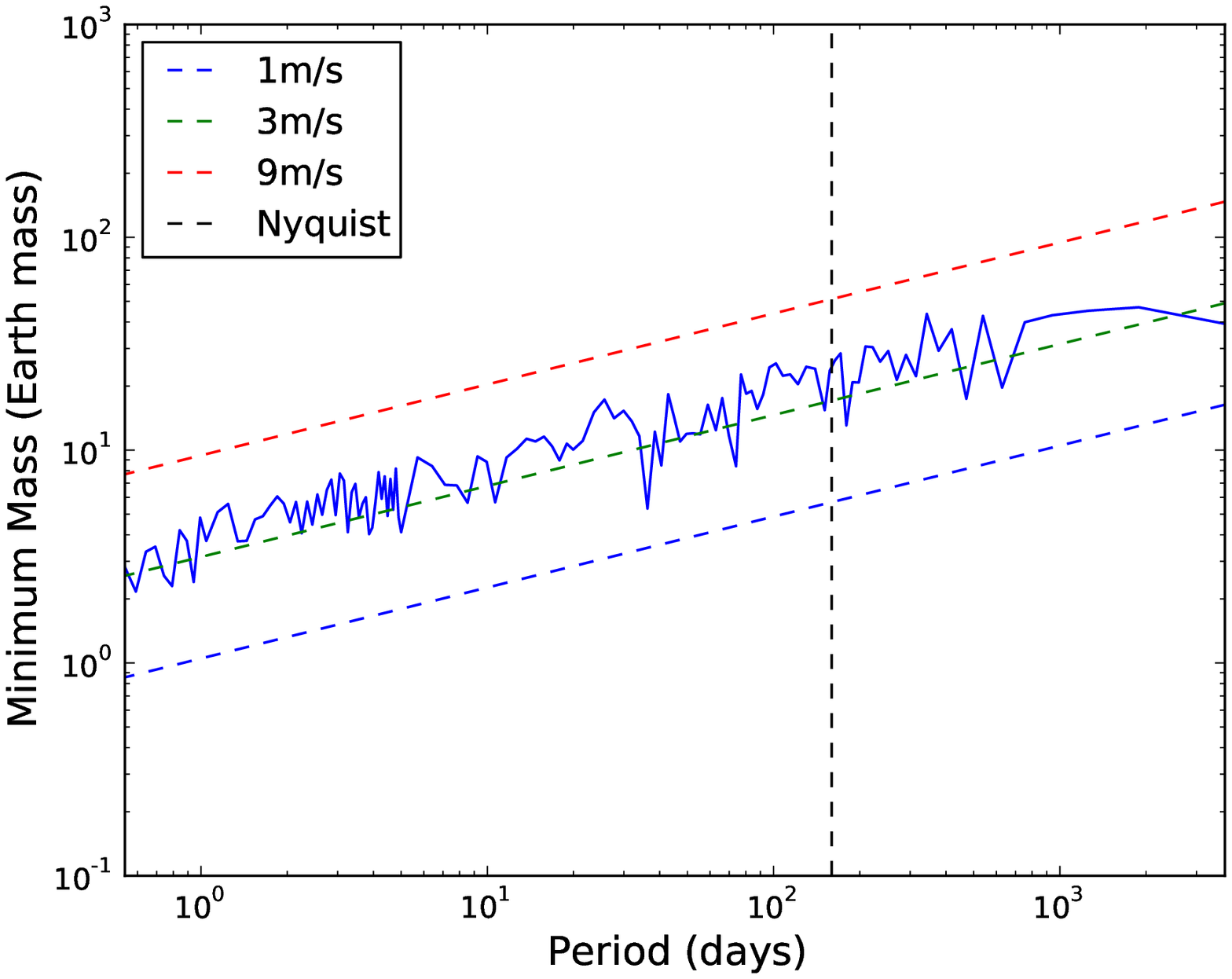}
\caption{Example of the GLS periodogram (top panel) and the detection limits (bottom panel) for the star \object{HD134440} with measurements from HARPS and KECK-HIRES. For both plots, a Keplerian fit was made. In the top panel, the power is plotted against the period. The horizontal solid line marks the power level for an FAP of $1\%$. The bottom panel plots the minimum planetary mass against the period. The solid line represents the detection limits for these data. The dashed lines indicate a circular planetary signal with an RV semi-amplitude of 1, 3 and 9 m s$^{-1}$ (lower to higher line).}
\label{FigEx}
\end{center}
\end{figure}

A frequency analysis of unevenly sampled data (like RV data) can be performed by using the GLS periodogram \citep{Sca82,Zech09}. The GLS is equivalent to a least-squares fitting of a full sine-wave, including weights and an offset, representing a circular orbit. In the resulting periodogram, the power $p(\omega)$ is calculated as a function of frequency. This power measures how much the fit to the measurements improves by using a sinusoid instead of a constant. This analysis can also be performed by replacing the ``sine-wave'' with a Keplerian function. An example of a GLS periodogram for a Keplerian fit is shown in Fig. \ref{FigEx} (top panel). This plot refers to a GLS of the RV data of \object{HD134440}, a star that has 16 measurements with an average rms of 4.9 m\,s$^{-1}$ and a timecoverage of 1885 days. We used both the circular and Keplerian approach.

The significance of a peak in the GLS periodogram can be determined analytically. However, not all values, like the number of independent frequencies, are properly determined in this way \citep{Zech09}. Alternatively, the significance can be accessed if we use a bootstrapping method \citep[e.g.][]{Endl01,Dum11}. Multiple time series of radial velocities are made by shuffling (with repetition) the real radial velocities while preserving the original times. On each virtual time series, a GLS is performed to determine the highest peak (frequency independent) in the periodogram. This can be used to determine the percentage of bootstrapped periodograms with maximum peaks above the one observed in the GLS of the actual data. This procedure allows us then to derive the false alarm probability (FAP) level (see top panel Fig. \ref{FigEx}). For this work, we have chosen to adopt $1000$ bootstrapping series to estimate the significance of the peaks.

Detection limits in RV data are derived by inserting a fake planetary signal in the data (circular or Keplerian). The procedure goes as follows. Virtual time series are made by adding these signals to the original data, which are treated as random noise. 
For a circular orbit, a fake signal

\begin{equation}
y(t) = K \sin\left[ \frac{2\pi}{P}t + \varphi\right] + c
\end{equation}

is added to the original data. Virtual series were made for periods $P$ from 0.5 to 3000 days, semi-amplitudes $K$ from 0 to 10 km s$^{-1}$ and ten phases $\varphi$, evenly separated by $\pi/10$. On each series, a GLS periodogram is performed. For each period, a signal is considered detected if the periodogram gives a peak at that period with a FAP of $1\%$ for all 10 phases. The minimum semi-amplitude $K$ for which a signal is detected expresses the lower limit for detectable planets in these data. 

The same approach can be taken for eccentric Keplerian signals. The fake signal, added to the original data, can in this case be described as follows:

\begin{equation}
y(t) =  a \cos\nu(t) + b \sin\nu(t) + c\qquad,
\end{equation}

with $a=K\cos\varpi$, $b=-K\sin\varpi$ and $c=K e \cos\varpi + \gamma$. Here, $K$ is the RV amplitude, $e$ the eccentricity, $\varpi$ the longitude of periastron, $\gamma$ the constant system RV and $\nu(t)$ the true anomaly. This true anomaly is a function of $t$, $e$, $P$ and $T_0$, the time of periastron passage. Again, virtual series were made for periods $P$ from 0.5 to 3000 days and semi-amplitudes $K$ from 0 to 10 km s$^{-1}$ were tried. For each period and semi-amplitude, 1000 virtual signals were created, with 10 different eccentricities $e$ (between 0 and 1), 10 different longitudes of periastron $\varpi$ (between 0 and $2\pi$) and 10 different times of periastron $T_0$ (between 0 and $P$), all evenly separated. In this case, a planet at a specific period $P$ is considered detected if the periodogram gives a peak with a FAP of $1\%$ for all 1000 signals.

The minimum semi-amplitudes can then be transformed into planetary masses (expressed in Earth mass) with the following formula:

\begin{equation}
M_p\sin i = 7.4\cdot 10^{-24} K\sqrt{1-e^2}\left( \frac{P M_{\ast}^2}{2\pi G}\right)^{1/3}\qquad,
\end{equation}

where the semi-amplitude $K$ is expressed in m/s, the period $P$ in days, the stellar mass $M_{\ast}$ in kg and the gravitational constant $G$ in m$^3$kg$^{-1}$s$^{-2}$. An example of these detection limits for \object{HD134440} is shown in Fig. \ref{FigEx} (bottom panel). A Keplerian signal was inserted to obtain these limits.

\begin{figure}[t!]
\begin{center}
\includegraphics[width=6.7cm,angle=270]{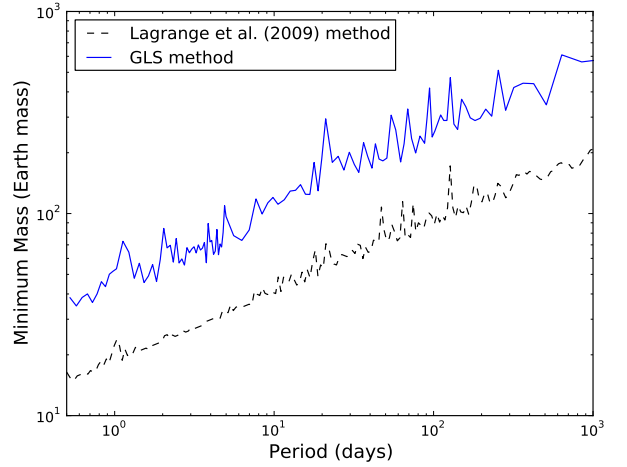}
\caption{Planetary mass is plotted against period for G19-27 (Keck sample). The blue (upper solid) line shows the detection limits for circular planetary signals in the data with a FAP of $1\%$. The black (lower dashed) line shows the detection limits based on the method described in \citet{Lag09}.}
\label{FigLag}
\end{center}
\end{figure}

As a comparison, for some stars, the detection limits were also calculated following the method described in \citet{Lag09}. Virtual RV sets are created with the expected RVs for a circular orbit, added with a random noise between $\pm$ RV error, where the RV error is the mean error of the real data. For every given period considered (between 0.5 and 1000 days), 200 virtual data sets were taken by varying the phase of the signal. This was then performed for different semi-amplitudes till the signal was detected. A signal was considered detected if the standard deviation of the real RV measurements was less than the average value of the virtual standard deviations. This resulted in detection limits with overall the same shape, but a factor of 2.5 lower, as can be seen in Fig. \ref{FigLag}. This shift can be explained by the fact that the periodogram analysis used in this work is more conservative in its definition of detectability, because it needs a peak above the $1\%$ FAP level for all phases. Because the overall shape is the same, the conservative periodogram analysis is favored for the purposes of this work.

Note that in all these analyses, a circular (resp. Keplerian) fit to RV data needs at least four (resp. seven) measurements. To be more conservative, the choice was made for at least six (resp. ten) measurements for the analysis.

\subsection{Stars with at least six measurements}

\begin{figure}[t!]
\begin{center}
\includegraphics[width=6.7cm,angle=270]{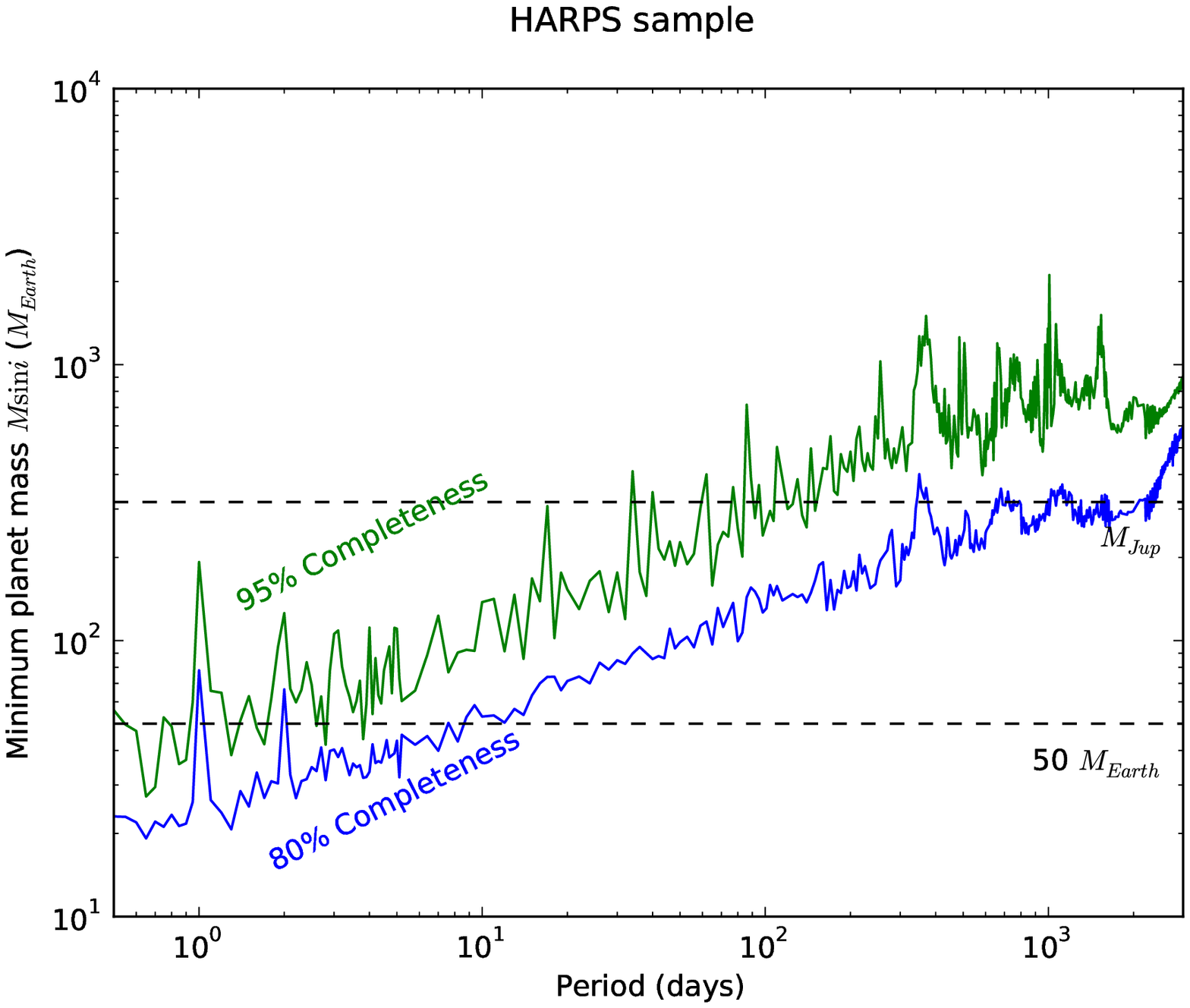}\\
\includegraphics[width=6.7cm,angle=270]{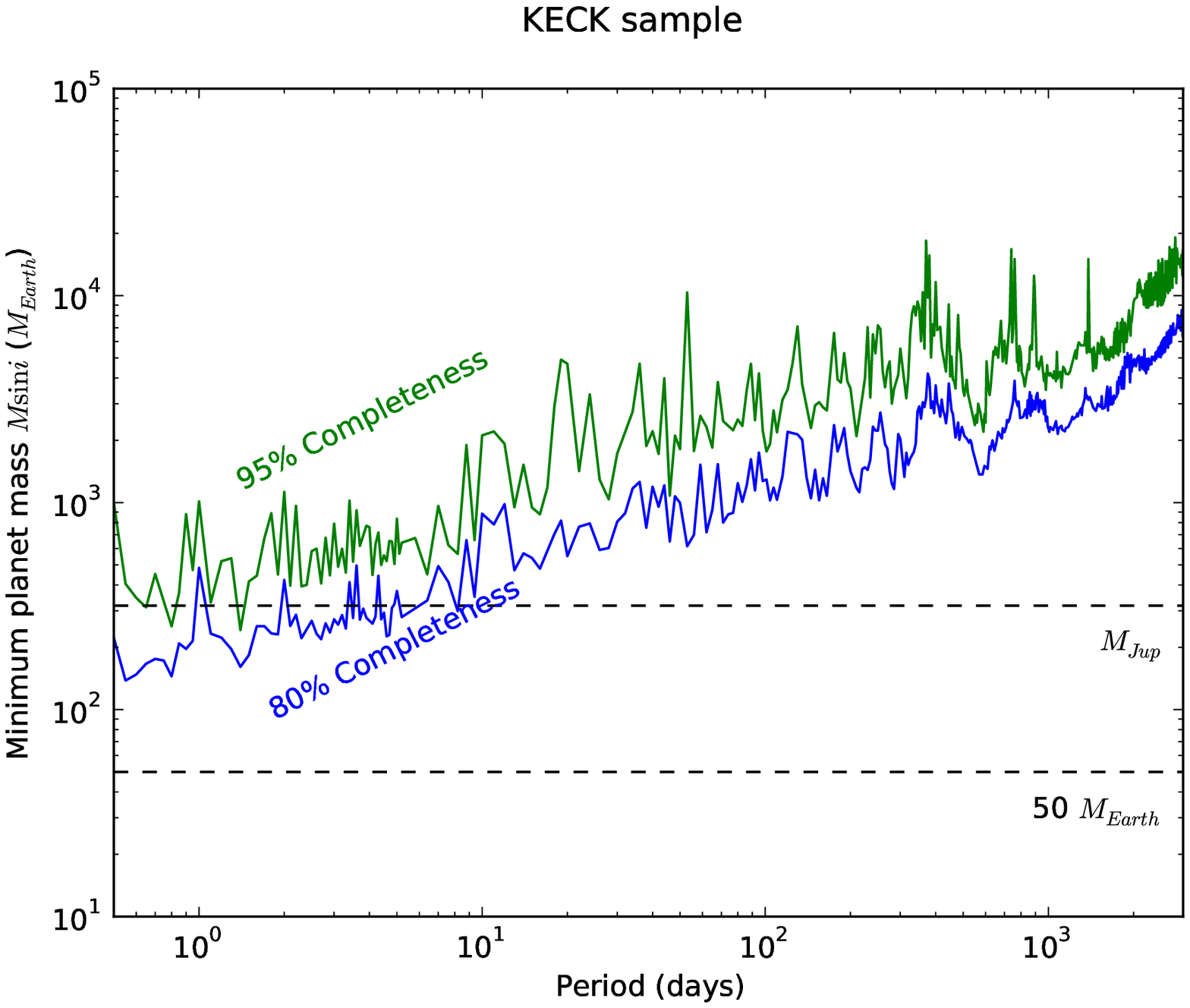}\\
\includegraphics[width=6.7cm,angle=270]{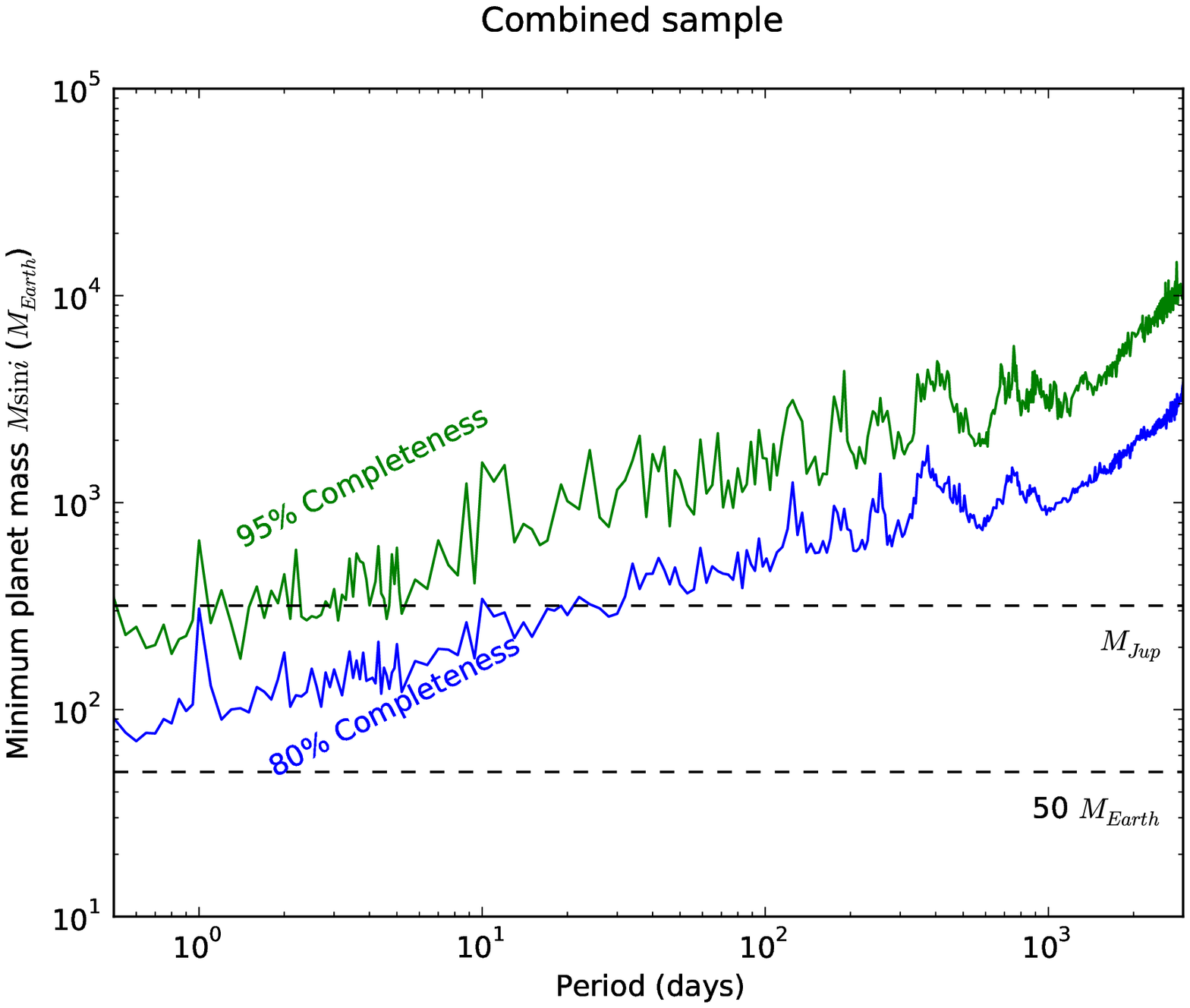}
\caption{Planetary mass is plotted against period. The blue line shows the detection limits for circular planetary signals in the data with a FAP of $1\%$. In the top panel, the data from the HARPS sample are shown. In the middle panel the data from the KECK-HIRES sample are shown and in the bottom panel, the combined sample is shown.}
\label{FigLimC}
\end{center}
\end{figure}

Detection limits based on a circular fit were made only for stars with at least six measurements. This resulted in 64, 50, and 114 stars ($72.7$, $31.25$, and $48.7\%$) for the HARPS, KECK-HIRES, and the combined sample, respectively.

The long-period signals (longer than the timespan of the measurements) produce a high-amplitude power in the GLS periodograms. To analyze the data for the presence of ``shorter'' period peaks, it is consequently necessary to remove these.
A linear function was fitted to the RV data of each star, using a least-squares fit. If the correlation coefficient $r^2$ was greater than $0.7$, the linear fit was considered relevant and subtracted from the original data. This was the case for \object{HD107094}, \object{HD11397}, \object{HD215257}, \object{HD123517}, \object{HD88725}, \object{HD144589}, and \object{HD113679} in the HARPS sample, \object{G135-46}, \object{G63-5}, \object{G197-45}, \object{HD134439}, \object{G237-84}, \object{HD192718}, \object{HD215257}, \object{HD7424}, and \object{G63-44} in the KECK-HIRES sample and all these objects together with \object{HD193901} in the combined sample. 
The planetary signals from the three confirmed planets in the HARPS sample \citep[with the parameters taken from][]{San11} were also subtracted, because we did not aim to confirm their existence but rather to search for additional signals.

Figure \ref{FigLimC} displays the detection limits for circular planetary signals in the three samples. Minimum planetary mass is plotted against period. The limits shown in the figure correspond to the values for which a signal can be detected in $80\%$ and $95\%$ (blue and green curve, respectively) of the stars in the sample. For clarity, we also include two dashed lines at Jupiter mass and at 50 M$_\oplus$, a value close to the generally accepted lower limit for giant planets. 
From these plots, it is clear that at a 80\% level no hot Jupiters (here defined as having an orbital period $<$10 days) could have been missed in the HARPS sample, though none was detected. In the KECK and the combined sample, most hot Jupiters should have been found.

\subsection{Stars with at least ten measurements}

\begin{figure}[t!]
\begin{center}
\includegraphics[width=6.7cm,angle=270]{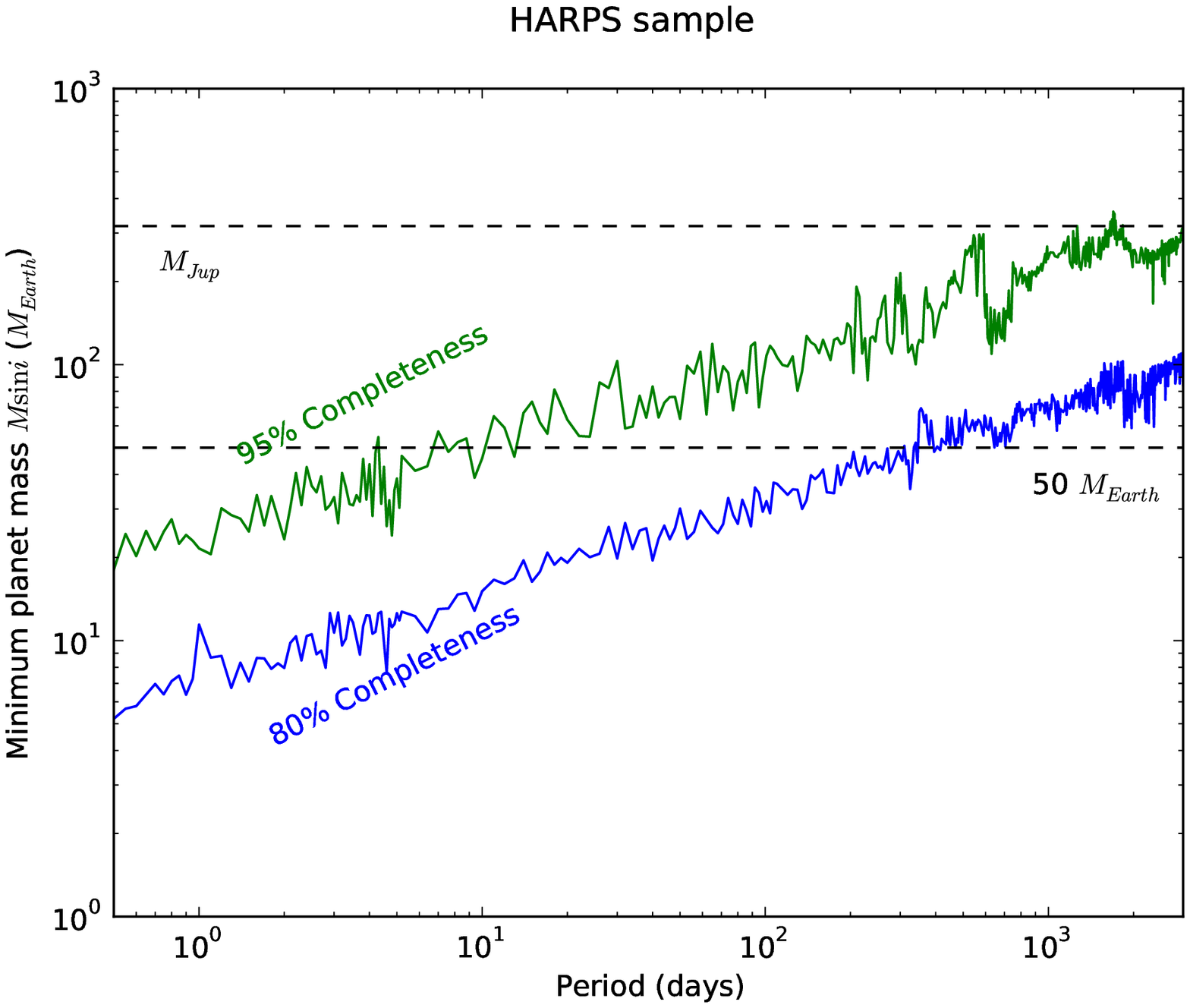}
\includegraphics[width=6.7cm,angle=270]{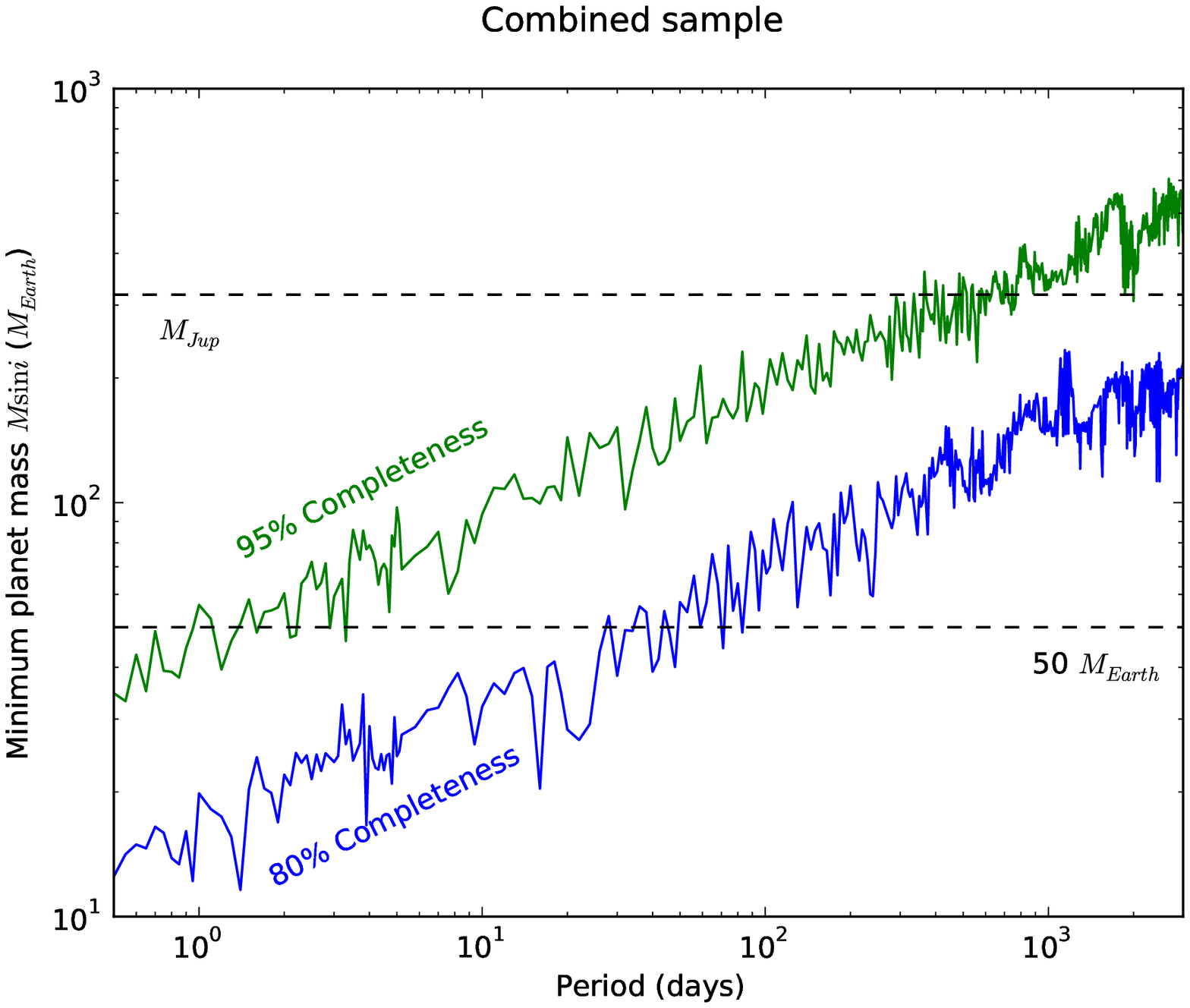}
\caption{Planetary mass is plotted against period. The blue line shows the detection limits for Keplerian (non-circular) planetary signals in the data with a FAP of $1\%$. In the top panel, the data from the HARPS sample are shown and in the bottom panel, the combined sample is shown.}
\label{FigLimK}
\end{center}
\end{figure}

For stars with at least ten measurements, the detection limits were also calculated for Keplerian signals. This was mainly useful for the HARPS sample, where more stars have dozens of measurements. In the HARPS, KECK-HIRES, and combined sample, respectively, there are 37, 7 and 47 stars ($42$, $4.4$ and $20.1\%$) with at least ten measurements.

Figure \ref{FigLimK} displays the Keplerian detection limits for the HARPS sample. The same trends and planetary signals as for the circular fit were subtracted before the analysis. The limits are for a $80\%$ and $95\%$ sample completeness (blue and green curve, respectively). The limits for the KECK sample are not shown because there are too few stars in the sample with at least ten measurements. Again, it is clear that no hot Jupiters were missed in the HARPS sample. The same is true for the combined sample. Note also that our data in the HARPS sample are sensitive to the detection of planets with masses above that of Jupiter for the whole covered period range.

%

\section{Planet frequency}\label{Freq}

With these limits and the number of planets found in the samples, a statistical analysis can be made of the giant planet frequency as a function of stellar metallicity. There are again several approaches to perform this analysis, among which the two main ones are a binning approach and a parametric approach (see below). The former makes use of a binomial distribution. The probability of finding $n$ detections in a sample of size $N$ can be calculated as a function of the true planet frequency $f_p$:

\begin{equation}
P(f_p;n,N)=\frac{N!}{n!(N-n)!}f_p^n(1-f_p)^{N-n}\qquad,
\end{equation}

This method is described in the appendix of \citet{Bur03}. For this asymmetric distribution, the errorbars can be computed by measuring the range in $f_p$ that covers $68\%$ of the integrated probability function. This is equivalent to the 1-sigma errorbars for a Gaussian distribution.

As seen above, no hot Jupiters were found in our samples, while the detection limits indicate that they most likely should have been detected. Zero (0) detections in a sample of 114 stars (stars in the combined sample with at least six measurements) leads to a frequency $f_p \leq 1.00\%$ (calculated from $f_p=0.37_{-0.4}^{+0.6}$). The frequency rises to $f_p \leq 2.36\%$ if only the 47 stars with at least ten measurements are taken into account. 

\begin{figure}[t!]
\begin{center}
\includegraphics[width=6.7cm,angle=270]{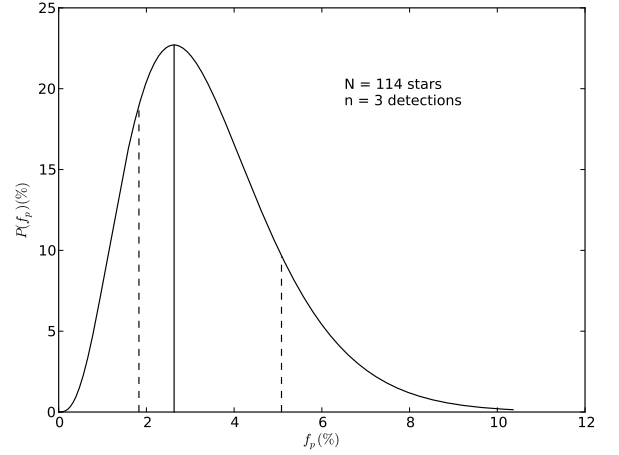}
\caption{Probability as a function of true planet frequency for a given amount of detections $n$ and sample size $N$. The solid vertical line denotes the observed planet frequency, while the dashed lines show the limits of the centered 68\% area, thus expressing the 1-sigma errorbars.}
\label{FigFreq}
\end{center}
\end{figure}

For giant planets in general (i.e. planets with a mass higher than $50 M_{Earth}$), there are three detections in a sample of 114 stars, which gives a frequency $f_p=2.63^{+2.5}_{-0.8}\%$ (see Fig. \ref{FigFreq}). In this sample of stars, $90\%$ have timespans longer than $900$ days. This makes these frequencies sensitive to planets with periods up to $1800$ days. If the sample is limited to only the 47 stars with at least ten measurements, the frequency becomes $f_p=6.38^{5.6}_{-2.0}\%$. For this smaller sample, this frequency is sensitive to planets with periods up to $2600$ days.

Our sample has a metallicity distribution that peaks around $-0.7$\,dex. If we divide it into two parts (above and below this limit), we find that on the high-metallicity side there are 67 (resp. 35) stars with at least six (resp. ten) measurements. This is also the side of the distribution of the detected giant planets. This thus leads to percentages of $f_p=4.48^{+4.04}_{-1.38}\%$ (resp. $f_p=8.57^{+7.21}_{-2.69}\%$). Around the stars in our sample with metallicities lower than $-0.7$ (47 stars with at least six measurements), no planets were detected. This again gives a frequency of $f_p \leq 2.36\%$. To check if the values are dependent on the choice of the bins, we repeated the calculation changing the position of the bins by 0.1dex (the typical 3-sigma error of the individual metallicity estimates), thus changing the high (low) metallicity side of the distribution to cover the range of [Fe/H]$>-$0.8\,dex ($\leq-$0.8\,dex, respectively). In these bins, the planet frequencies for stars with at least six measurements become $f_p=3.70^{+3.39}_{-1.13}\%$ resp. $f_p\leq 3.32\%$. Both results are comparable within the errorbars.

Alternatively, a parametric approach, similar to the one used by \citet{John10}, was also considered. Here, the data are fitted with a functional form, dependent on stellar metallicity. As a functional form, we chose

\begin{equation}
f_p=C\cdot 10^{\beta [Fe/H]}\qquad,
\end{equation}

which is typically used for solar neighbourhood samples. The best parameters $(\beta,C)$ are then determined by using a numerical fitting procedure, based on Bayesian inference. Details of this procedure can be found in \citet{John10}. As a prior, the choice was made for uniformly distributed parameters over $[0.0,3.0]$ and $[0.01,0.30]$ for $\beta$ and $C$, respectively. The best parameters were found to be $\beta = 1.3$ and $C = 0.17$. The mean metallicity for stars with six measurements and a metallicity higher than $-0.7$\,dex, is $-0.55$\,dex. For the derived $\beta$ and $C$, the corresponding expected frequency would be $3.29\%$, compatible with the results of the binning procedure. Owing to the limited size of the sample and because we are exploring a metallicity regime not previously explored with high-precision radial velocities, it is unclear which appropriate functional form should be used.

\section{Conclusions and discussion}\label{Con}

Radial velocities of two samples of metal-poor solar-type stars, taken with two different instruments (HARPS and KECK-HIRES), were used to detect extrasolar planets. Only three giant long-period planets were found out of the 234 stars, together with one giant candidate. Fourteen stars were present in both samples, but the expanded datasets and extended baseline of the observations did not reveal any additional signals. 

After subtracting linear trends or planetary signals, detection limits in the samples were calculated. The method was based on a GLS periodogram analysis and bootstrapping. Limits were calculated for circular and Keplerian signals. These lower limits, as shown in Figs. \ref{FigLimC} and \ref{FigLimK}, are expressed in minimum planetary mass and period. For the stars in the KECK sample, detection limits were already derived by \citet{Soz09}. They used a method based on $\chi^2$- and $F$-tests. The detection limits derived in this work perfectly agree with their results.

A statistical analysis was performed to estimate the planet frequency around metal-poor main-sequence stars. Taking into account only the stars with at least six measurements, we showed that the frequency of hot Jupiters around metal-poor stars is lower than $1.00\%$. This is consistent with previous studies \citep{Udry07}.

Giant planets, however, seem to be more frequent around these stars. The detection limits show that most of the giant planets should have been detected in this sample. A frequency of $2.63\%$ for giant planets around metal-poor stars was calculated for stars with at least six measurements, with a sensitivity to periods up to $1800$ days. If indeed a giant planet was missed in the sample, the frequency would be even higher. According to several studies \citep[e.g.][]{San04,Fis05}, $3\%$ is the frequency of giant planets around stars of solar metallicity. Given the same number, derived in this work, for metal-poor stars, this can mean two things: either the planet frequency becomes constant for stars with $[Fe/H]\leq 0.0$ \citep[for a discussion, see also e.g.][]{San04,Udry07,John10} or previous frequency-models should be higher.

The metallicities of the stars with discovered giant planets all lie above $-0.7$\,dex. Within this metallicity bin, the planet frequency increases to $4.48^{+4.04}_{-1.38}\%$. 
For a metallicity of $-0.55$ dex and a stellar mass of $0.8M_{\odot}$ (mean value for this sample), previous studies report values of $f_p = 1.22^{+0.7}_{-0.5}\%$ \citep{John10} and $f_p = 0.14\%$ \citep{Sou11b}. However, in both cases, a powerlaw was fitted over the whole metallicity range (up to $0.6$ and $0.5$ dex, respectively). In the low-metallicity end, their fit is clearly lower than their observed fraction ($\sim 4.6\%$ and $3.77\%$, respectively). The value reported here is thus higher than previous fits, but consistent within one sigma with the observed fraction in these previous studies.

For stars with metallicities lower than $-0.7$, the frequency is lower than $2.36\%$. In this context, it is worth mentioning that so far, only one (giant) planet has been detected around a main-sequence star with metallicity lower than $-0.6$ dex \citep{Coc07}. 
There are some other candidates where planets orbit stars with metallicities lower than $-0.6$ dex \citep{Nie09,Set03,Set10}. These stars are all giants, however. Furthermore, there is a candidate planet detected by imaging that orbits a young main-sequence star with a metallicity of $-0.64$ dex \citep{Cha05}. These planets are therefore not relevant for the purpose of this work.

All the above results are strong evidence that giant planet frequency is a non-constant function of stellar metallicity. This was already established for the high-metallicity tail, but this work shows that it is also true for lower metallicities. Moreover, the frequencies are probably higher than previously thought. A powerlaw may not be the best function to describe the planet frequency over the whole metallicity range. As mentioned before, this correlation between giant planet frequency and stellar metallicity favors the core-accretion model as the main mechanism for giant planet formation. 

With the statistics presented in this work, a metallicity limit can be established below which no giant planets can be found anymore. According to these statistics, this metallicity limit would be about $-0.5$ - $-0.6$ dex. This value agrees with the recent results of the theoretical study of \citet{Mor12}. These authors looked at correlations between stellar and planetary properties, based on a synthetic planet population, built by the core accretion model. They found that giant planets are not formed below $-0.5$ dex.

However, more data are still needed to produce better statistics. Future missions, such as Gaia, will produce better precision in the data. For example, in its all-sky global astrometric survey, Gaia will probe thousands of nearby metal-poor stars for gas giant planets within 3-4 AU \citep[e.g.][]{Soz10}, thus crucially helping to shed light on the planet-metallicity connection.

\begin{acknowledgements}

We thank the ananymous referee for his/her useful comments.
This work was supported by the European Research Council/European Community under the FP7 through Starting Grant agreement number 239953. NCS also acknowledges the support from Funda\c{c}\~ao para a Ci\^encia e a Tecnologia (FCT) through program Ci\^encia\,2007 funded by FCT/MCTES (Portugal) and POPH/FSE (EC), and in the form of grants reference PTDC/CTE-AST/098528/2008 and PTDC/CTE-AST/09860/2008.

\end{acknowledgements}

\bibliographystyle{aa} 
\bibliography{References.bib}

\end{document}